# ICLR Points: How Many ICLR Publications Is One Paper in Each Area?

Zhongtang Luo
Purdue University
West Lafayette, Indiana, USA
luo401@purdue.edu

## ABSTRACT

Scientific publications significantly impact academic-related decisions in computer science, where top-tier conferences are particularly influential. However, efforts required to produce a publication differ drastically across various subfields. While existing citation-based studies compare venues within areas, cross-area comparisons remain challenging due to differing publication volumes and citation practices.

To address this gap, we introduce the concept of ICLR points, defined as the average effort required to produce one publication at top-tier machine learning conferences such as ICLR, ICML, and NeurIPS. Leveraging comprehensive publication data from DBLP (2019–2023) and faculty information from CSRankings, we quantitatively measure and compare the average publication effort across 27 computer science subareas. Our analysis reveals significant differences in average publication effort, validating anecdotal perceptions: systems conferences generally require more effort per publication than AI conferences.

We further demonstrate the utility of the ICLR points metric by evaluating publication records of universities, current faculties and recent faculty candidates. Our findings highlight how using this metric enables more meaningful cross-area comparisons in academic evaluation processes. Lastly, we discuss the metric's limitations and caution against its misuse, emphasizing the necessity of holistic assessment criteria beyond publication metrics alone.

## 1 INTRODUCTION

Scientific publications are critical to disseminate study results and facilitate academic discussions. As a result, publication records in prestigious venues are often used as criteria for faculty hiring, promotion and tenure cases [4].

Computer science differs from other subjects in that the field puts a greater emphasis on conference publications compared to journal publications [5]. Previous researches have suggested that conference papers in computer science have a higher average two-year citation count than journal papers [3]. Specifically, top-tier conferences have the highest average citation rate among comparable venues [12]. Therefore, *CSRankings*, one widely-used CS GOTO (Good data,

Open, Transparent, and Objective) ranking system [2, 11], uses exclusively the number of top-tier conference publications as the criteria to determine research output [1].

However, the relative effort to publish a paper in each area of computer science are not the same. For example, in 2024, based on data from CSRankings, top-tier conferences in machine learning (ICLR, ICML and NeurIPS) have seen 9364 publications, whereas comparable conferences in algorithms & complexity (FOCS, SODA and STOC) have only 191 publications [1]. Such a drastic difference raises a natural question: is it possible to quantitatively determine the average efforts to produce one publication in each area's top-tier conferences?

Existing studies on publication records focus on citation analysis of publications [6, 8–10]. Such studies are not meaningful when comparing across areas. On the contrary, in areas where the average effort is lower, citation counts tend to be higher due to the larger number of publications.

In this paper, we motivate the concept of an *ICLR point*, the average effort to produce one publication in the top-tier conferences in machine learning such as ICLR. We then use this concept to measure the average effort to produce one publication in each area's top-tier conferences. It is worth noting that the effort spent in one publication is influenced by many factors, such as the difficulty of the field, number of grants and graduate students, novelty and completeness requirement of the conferences. Therefore, the result only indicates the average effort to produce one publication, and does not reflect the inherent difficulty of the areas or the merit of the publications/areas.

After that, we perform analysis on the results by measuring universities', existing faculties' and faculty candidates' publications using ICLR points. Our analysis shows some interesting facts about comparing universities, faculties and candidates across areas, which we believe is only possible through the use of a cross-area metric such as ICLR points.

Finally, we give a discussion on the limitations of the ICLR points metric and its implications for faculty hiring, promotion and tenure cases. We believe these insights can help the computer science community better understand the publication practices across different areas and improve the evaluation of faculty candidates.

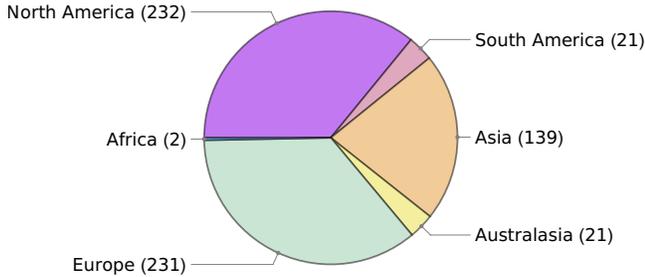

North America (232)

South America (21)

Africa (2)

Asia (139)

Australasia (21)

Europe (231)

**Figure 1: Distribution of universities in the dataset by continent.**

## 2 SCOPE & DATASETS

In this section, we address the scope and datasets we use for our analysis.

### 2.1 Areas & Conferences

We note that the definition of area and top-tier conferences in the area is subjective. For the purpose of our analysis, we use the areas and conferences defined by CSRankings.

According to CSRankings, the areas are chosen according to the following rationale [1].

- Areas are either based on research-focused ACM SIGs or most established research-centric areas.
- At least 50 R1 institutions must have publications in the top conferences in the area in the last 10 years.

The conferences listed were developed in consultation with faculty across a range of institutions and are generally uncontroversial [1]. A list of areas and conferences is shown in Table 3 in Appendix B. The list includes 4 parent areas, 27 sub-areas and 64 top-tier conferences.

### 2.2 Publications

We use the DBLP computer science bibliography [7] to obtain the publication data for the top-tier conferences in each area. DBLP has maintained a relatively complete and accurate record of computer science publications, including disambiguation of authors with the same names, and is the data source for CSRankings as well.

### 2.3 Faculties & Universities

CSRankings define a faculty as anyone who is a full-time (at least 75% appointment), tenure-track faculty member on a given campus who can solely advise PhD students in computer science [1].

Based on the data, we identified 30,402 faculties from 646 universities across the world, including 199 United States institutions.

While the data is not an exhaustive list of all faculties and universities, we believe it is a representative sample of the computer science research community. The distribution of universities by continent is shown in Figure 1.

## 3 METHODOLOGY

We aim to quantify the average effort required to produce a single publication in each research area's top-tier conferences, expressed in units of *ICLR points*. One ICLR point is defined as the average research effort needed to produce one publication in a premier machine learning conference, such as ICLR. We use this naming convention because ICLR is widely recognized as a representative and influential top-tier venue in machine learning.

By definition,

$$(\text{Average effort}) = \frac{(\text{Total effort})}{(\text{Total number of publications})}.$$

Our task therefore reduces to estimating both the total research effort and the total number of publications in each area. We make the assumptions below to simplify the estimation process.

**Assumption 1.** *An average faculty in each field spends the same amount of research effort per year.*

This allows us to approximate total effort in a field by counting its affiliated faculty. Although individual research productivity varies, we assume that across a sufficiently large population and time window, these variations average out.

**Assumption 2.** *Most academic research effort is directed toward publications in top-tier conferences.*

This assumption enables us to treat the number of top-tier publications as a reasonable proxy for productive research output. While other outputs (e.g., journal papers, patents, workshops) also matter, top-tier conference publications remain the dominant metric in many computer science subfields.

### 3.1 Total Effort

Directly measuring research effort is challenging. Instead, we use faculty count as a proxy. Under Assumption 1, the total effort in a field is proportional to the number of faculty members contributing to that field.

To account for interdisciplinary researchers, we apply fractional attribution: if a faculty member publishes in multiple areas, we divide their effort evenly across those areas. For example, a researcher publishing in both computer security and cryptography and is assumed to contribute 0.5 effort to each.

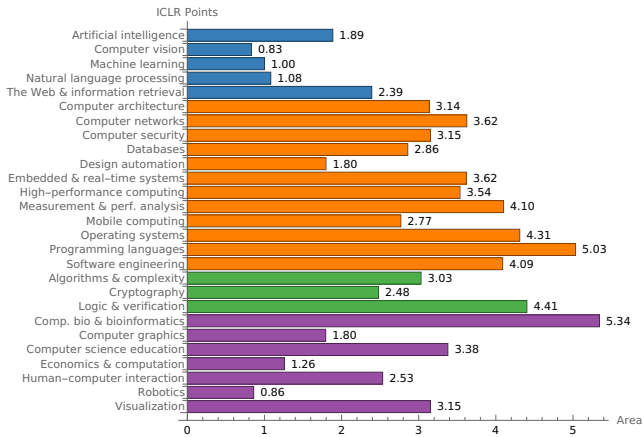

**Figure 2: Average effort to produce one publication in each area's top-tier conferences in the unit of ICLR points. We note that the effort spent in one publication is influenced by many factors, such as the difficulty of the field, number of grants and graduate students, novelty and completeness requirement of the conferences. Therefore, the result only indicates the average effort to produce one publication, and does not reflect the inherent difficulty of the areas or the merit of the publications/areas.**

## 3.2 Total Number of Publications

We define total publications in a field as the number of papers published in its top-tier conferences over a 5-year window (2019–2023). This balances recentness with statistical stability. We extract publication data from DBLP [7], which comprehensively indexes all target conferences during this period.

## 4 RESULTS

Based on the datasets and methodology described above, we compute the average effort to produce one publication in each area's top-tier conferences in the unit of ICLR points. The results are shown in Figure 2, and the full data is available in Appendix B.

We observe that the results match our anecdotal experience.

(1) Conferences in machine learning, computer vision, and natural language processing have similar average efforts to produce one publication.

(2) Conferences in systems generally have a higher average effort to produce one publication than conferences in AI.

(3) Conferences in "traditional" system areas (computer networks, operating systems, and programming languages) have a higher average effort to produce one publication than security conferences (computer security and cryptography).

## 5 ANALYSIS OF THE RESULTS

In this section, we analyze the measure of ICLR points under different scenarios to understand its limitations and implications.

## 5.1 Measurement over All Universities

We measure the ICLR points of all universities in the dataset based on their faculties' publications over two time periods: (1) 1970–2024 (all available data) and (2) 2019–2023, using adjusted ICLR points — defined as rewarding each faculty with each publication's ICLR points divided by the number of authors — as the metric. We compare the ranking results with the ranking from CSRankings over the same time periods. The results are in Table 1.

We observe that ranking by ICLR points is largely consistent with the ranking by CSRankings at the top level: the top 10 universities in ICLR points are also in the top 10 in CSRankings during 1970–2024.

On the other hand, discrepancies between rankings highlight an interesting comparison between different ranking metrics. We give an area-by-area radar graph of

(1) the two top institutions (Massachusetts Institute of Technology & Univ. of Illinois at Urbana-Champaign) that have different rankings during 1970–2024,

(2) the two institutions (Peking University (+5) & Princeton University (+12)) that gain significant ranking by ICLR points during 1970–2024,

(3) the two institutions (Zhejiang University (+8) and KAIST (+5)) that gain significant ranking by ICLR points compared to CSRankings during 2019–2023, and

(4) the two institutions (Northeastern University (-10) and University of Texas at Austin (-4)) that lose significant ranking by ICLR points compared to CSRankings. Incidentally, they also lose significant ranking in CSRankings (-6 and -9 respectively) during 2019–2023.

in the top 20 universities in Figure 3 with data in the respective time period.

The figure shows some interesting trends between the two metrics.

(1) Universities that gain ranking in ICLR points compared to CSRankings tend to specialize in one field. Specifically, Massachusetts Institute of Technology and Princeton University all have a large number of publications in algorithms & complexity, while Peking University, Zhejiang University and KAIST all have a large number of publications in AI.

**Table 1: Ranking of universities by ICLR points from a time period of (1) left: 1970–2024 (all available data) and (2) right: 2019–2023. This table lists the top 20 schools in both the ICLR points and the CSRankings ranking, in descending order of ICLR points. The rank in CSRankings is shown in parentheses. We note that CSRankings sometimes has ties in the ranking.**

| Rank (CSRankings) | School | ICLR Points |
|---|---|---|
| 1 (1) | Carnegie Mellon University | 4628.37 |
| 2 (3) | Massachusetts Institute of Technology | 3179.18 |
| 3 (4) | Univ. of California - Berkeley | 2866.72 |
| 4 (2) | Univ. of Illinois at Urbana-Champaign | 2732.59 |
| 5 (5) | Univ. of California - San Diego | 2527.72 |
| 6 (6) | Stanford University | 2254.38 |
| 7 (10) | University of Washington | 2157.97 |
| 8 (7) | University of Michigan | 2151.87 |
| 9 (9) | Cornell University | 2149.08 |
| 10 (8) | Georgia Institute of Technology | 2081.84 |
| 11 (13) | Tsinghua University | 1862.28 |
| 12 (11) | ETH Zurich | 1716.06 |
| 13 (12) | University of Maryland - College Park | 1661.12 |
| 14 (14) | University of Toronto | 1598.75 |
| 15 (20) | Peking University | 1585.11 |
| 16 (16) | University of Pennsylvania | 1535.81 |
| 17 (29) | Princeton University | 1514.10 |
| 18 (21) | New York University | 1498.43 |
| 19 (25) | Technion | 1492.40 |
| 20 (15) | Columbia University | 1474.88 |
| 21 (17) | University of Texas at Austin | 1467.08 |
| 22 (19) | National University of Singapore | 1452.02 |
| 28 (18) | Northeastern University | 1314.74 |

| Rank (CSRankings) | School | ICLR Points |
|---|---|---|
| 1 (1) | Carnegie Mellon University | 1222.61 |
| 2 (3) | Tsinghua University | 821.70 |
| 3 (2) | Univ. of Illinois at Urbana-Champaign | 754.56 |
| 4 (6) | Peking University | 753.59 |
| 5 (8) | Massachusetts Institute of Technology | 698.23 |
| 6 (4) | Univ. of California - San Diego | 679.11 |
| 7 (6) | Shanghai Jiao Tong University | 673.44 |
| 8 (5) | Georgia Institute of Technology | 632.33 |
| 9 (17) | Zhejiang University | 598.48 |
| 10 (14) | Univ. of California - Berkeley | 582.98 |
| 11 (15) | Stanford University | 557.84 |
| 12 (11) | University of Washington | 543.65 |
| 13 (18) | KAIST | 543.62 |
| 14 (8) | University of Michigan | 541.40 |
| 15 (10) | ETH Zurich | 528.23 |
| 16 (11) | University of Maryland - College Park | 486.90 |
| 17 (21) | HKUST | 480.50 |
| 18 (11) | Cornell University | 475.72 |
| 19 (15) | University of Toronto | 475.11 |
| 20 (26) | Chinese Academy of Sciences | 467.97 |
| 21 (19) | National University of Singapore | 450.70 |
| 25 (19) | Northeastern University | 408.76 |

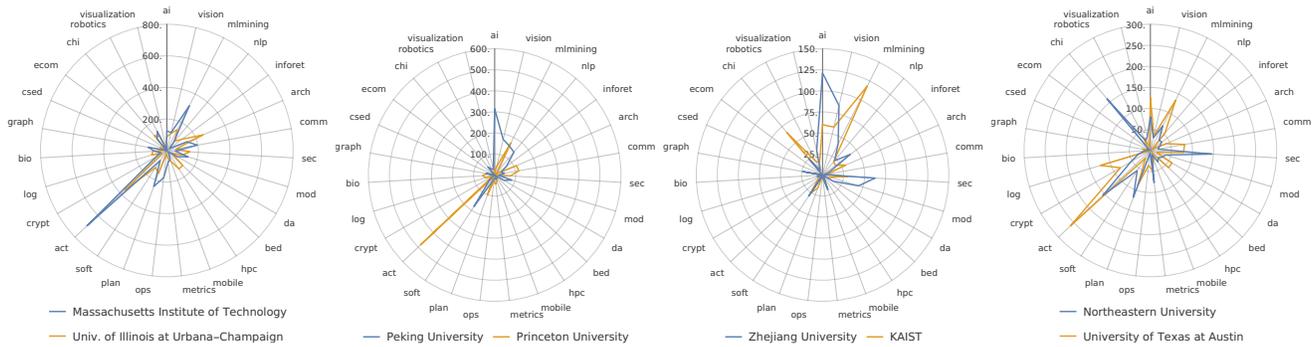

**Figure 3: Radar graph of (1) Massachusetts Institute of Technology and Univ. of Illinois at Urbana-Champaign during 1970–2024, (2) Peking University and Princeton University during 1970–2024, (3) Zhejiang University and KAIST during 2019–2023, and (4) Northeastern University and University of Texas at Austin during 1970–2024. The radar graph shows the distribution of ICLR points across areas for the universities in the dataset.**

For example, Massachusetts Institute of Technology has a combined point of 695.29 in algorithms & complexity, which is almost 2 times compared to Univ. of Illinois at Urbana-Champaign with 364.39. Princeton University also have most points in algorithms & complexity, more than 4 times compared to their second area.

(2) Universities that lose ranking in ICLR points tend to have a more balanced distribution of publications across areas.

For example, Northeastern University have a balanced distribution of publications across areas, with no particularly notable areas. University of Texas at Austin have a large number of publications in theory, but also have a significant number of publications in other areas.

We conclude that this difference in metrics can be largely attributed to CSRankings' ranking formula. Notably, while CSRankings also counts each publication once, with credit adjusted by splitting evenly across all co-authors, it uses the

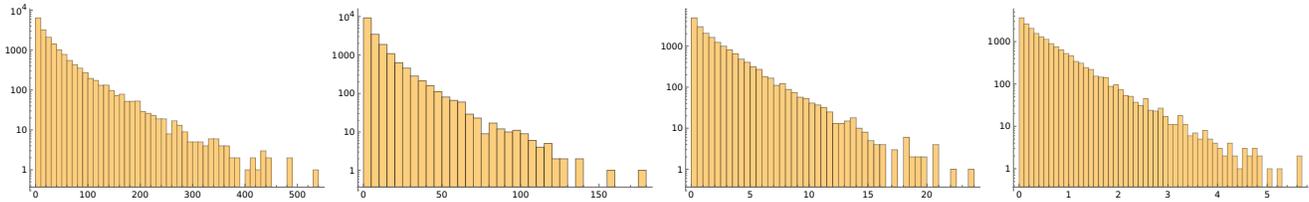

Figure 4: Distribution of sum of ICLR points, sum of adjusted ICLR points, average ICLR points per year, and average adjusted ICLR points per year for faculties in the dataset. Note the y-axis is in logarithmic scale.

geometric mean of the adjusted counts across every area

$$\text{(Average count)} = \sqrt[N]{\prod_{i=1}^{N}((\text{Adjusted counts})_i + 1)}$$

as the basis of its ranking [1]. Therefore, a hypothetical school publishing 10 papers in area A and 10 papers in area B will be ranked higher than a school publishing 50 papers in area A and no papers in area B, as well as a school publishing no papers in area A and 50 papers in area B.

We recognize that rankings are subjective and that different metrics yielding different results may not be better or worse than the other. However, we believe that the introduction of ICLR points highlights the incredible amount of work in algorithms & complexity and artificial intelligence done by some universities, and provides an interesting and meaningful metric for comparing publication efforts across areas.

## 5.2 Measurement over All Faculties

For the faculties in the dataset, we measure the sum of ICLR points for their publications. We also measure the sum of adjusted ICLR points for their publications, defined as each publication's ICLR points divided by the number of authors. Furthermore, we measure the average (adjusted) ICLR points per year, defined as the sum of (adjusted) ICLR points divided by the number of years since the faculty's first publication. The results are shown in Figure 4.

Based on the figure, we see that all distributions roughly follow an exponential distribution, except for a few outliers with high ICLR points. We list the top 5 faculties in each metric in Table 2.

We learn a few interesting facts from the table.

(1) Faculties with high ICLR points tend to work in multiple areas. Top 5 faculties in total ICLR points and in total adjusted ICLR points all work in no less than 5 areas.

(2) Faculties with high average adjusted ICLR points per year tend to work in theory areas. For example, 4 out of the top 5 faculties in average adjusted ICLR points per year (except Heng Huang, who mostly works in AI) has published in an algorithms & complexity conference.

## 5.3 Measurement of Faculty Candidates

We sample the recent 57 PhD/postdocs that came to a top-30 computer science department (according to CSRankings) and gave a job talk in 2024–2025.

We measure the (adjusted) ICLR points according to their publications. Additionally, we measure the ICLR points using their first-authored publications. For conferences in the parent area of theory, we assume the authors are in alphabetical area and allocate the adjusted ICLR points to the candidates regardless of the order. The results are shown in Figure 5.

We observe that the average ICLR points for first author publications are around 7.5, which is in line with our anecdotal intuition about an exceptional candidate.

Furthermore, the distribution in all three figures roughly follows a normal distribution. We see that both the AI and the system area candidates also follows a rough normal distribution, though the system candidates have a significantly higher average. We think this is in line with the recent department effort on hiring AI talents.

We also notice a few limitations of the ICLR points metric.

(1) The metric only measures top-tier conference publications, which may not be representative of the candidates' overall publication records. For instance, one candidate has multiple publications in COLT, which is not listed in CSRankings.

(2) The metric does not account for diverse experiences and contributions of candidates. Some strong candidates also have extensive industry experience, which is not reflected in their publication records.

## 6 RELATED WORK

Most existing studies on publication records focus on the citation analysis of publications [6, 8–10]. We consider such study inadequate when comparing publications across areas as citation counts are heavily influenced by the number of publications in the area. For instance, according to Google Scholar in 2025, NeurIPS has an h5-index of 337, while STOC has an h5-index of 57. It is unrealistic to claim that NeurIPS papers are 6 times more impactful than STOC papers.

**Table 2: Top 5 faculties in the dataset based on different metrics. Columns: (1) Name, (2) number of areas, (3) year of first publication, (4) total ICLR points, (5) total adjusted ICLR points, (6) average ICLR points per year, and (7) average adjusted ICLR points per year.**

| Total ICLR Points | | | | | | |
|---|---|---|---|---|---|---|
| Name | # | Year | Points | Adjusted | Avg. Pts | Avg. Adj |
| Ion Stoica | 15 | 1996 | 538.55 | 103.42 | 18.57 | 3.57 |
| Onur Mutlu | 12 | 2003 | 489.37 | 86.88 | 22.24 | 3.95 |
| Mahmut Kandemir | 11 | 1997 | 486.35 | 119.21 | 17.37 | 4.26 |
| Scott Shenker | 11 | 1987 | 443.30 | 105.89 | 11.67 | 2.79 |
| Moshe Vardi | 8 | 1982 | 441.13 | 176.77 | 10.26 | 4.11 |

| Adjusted ICLR Points | | | | | | |
|---|---|---|---|---|---|---|
| Name | # | Year | Points | Adjusted | Avg. Pts | Avg. Adj |
| Moshe Vardi | 8 | 1982 | 441.13 | 176.77 | 10.26 | 4.11 |
| Kang Shin | 11 | 1980 | 414.21 | 159.05 | 9.20 | 3.53 |
| Mikkel Thorup | 5 | 1994 | 278.64 | 139.46 | 8.99 | 4.50 |
| Thomas Henzinger | 11 | 1989 | 438.07 | 137.52 | 12.17 | 3.82 |
| David Woodruff | 10 | 2002 | 343.48 | 128.99 | 14.93 | 5.61 |

| Average ICLR Points Per Year | | | | | | |
|---|---|---|---|---|---|---|
| Name | # | Year | Points | Adjusted | Avg. Pts | Avg. Adj |
| Yang Liu | 11 | 2009 | 379.45 | 59.63 | 23.72 | 3.73 |
| Onur Mutlu | 12 | 2003 | 489.37 | 86.88 | 22.24 | 3.95 |
| Yong Li | 8 | 2015 | 209.71 | 39.74 | 20.97 | 3.97 |
| Dacheng Tao | 7 | 2004 | 439.32 | 98.39 | 20.92 | 4.68 |
| Sergey Levine | 5 | 2009 | 330.39 | 74.37 | 20.65 | 4.65 |

| Average Adjusted ICLR Points Per Year | | | | | | |
|---|---|---|---|---|---|---|
| Name | # | Year | Points | Adjusted | Avg. Pts | Avg. Adj |
| Heng Huang | 7 | 2008 | 337.41 | 96.21 | 19.85 | 5.66 |
| David Woodruff | 10 | 2002 | 343.48 | 128.99 | 14.93 | 5.61 |
| Jason Li | 3 | 2018 | 97.08 | 36.56 | 13.87 | 5.22 |
| K. Kawarabayashi | 6 | 2005 | 245.63 | 100.08 | 12.28 | 5.00 |
| Aviad Rubinstein | 3 | 2014 | 125.83 | 53.24 | 11.44 | 4.84 |

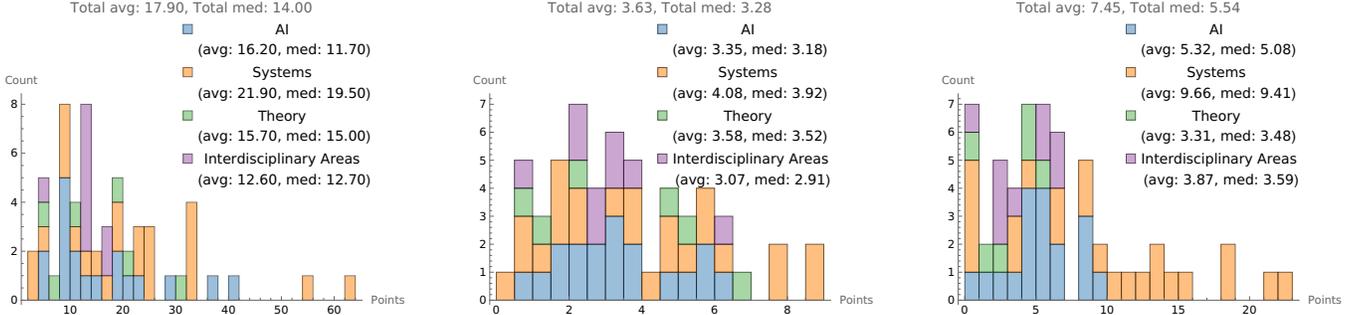

**Figure 5: Distribution of (1) total ICLR points, (2) total adjusted ICLR points, and (3) total ICLR points based on first-authored publications for faculty candidates in the dataset.**

In contrast, our study aims to bring some perspective into comparison of publications across areas by measuring the average effort to produce one publication in each area's top-tier conferences. We believe this is a more meaningful metric than citation counts when comparing publications across areas.

## 7 LIMITATIONS

We observe that while ICLR points provide an interesting and important perspective on comparing publications across areas, it has its limitations in several aspects.

**Subjectivity of Areas and Conferences** While this study follows the areas and conferences defined by CSRankings, the definition of areas and top-tier conferences is still subjective. For example, CSRankings does not include COLT as a top-tier conference in machine learning. As a result, ICLR points may not accurately reflect the effort to produce one publication in machine learning, if much effort is spent on publications in COLT instead.

**Interdisciplinary Areas** Interdisciplinary areas often have practices that align with other subjects more than computer science. ICLR points often fail to capture these aspects. For example, in bioinformatics and education, the practice of publishing in journals is more common than in conferences. As a result, the ICLR points metric may not accurately reflect the effort to produce one publication in these areas, since the low amount of publications in conferences may be due to the fact that more efforts are spent for journals instead.

**Potential for Misuse** While we believe that ICLR points provide a meaningful metric for comparing publications across areas, it is important to note that the metric can be misused. For example, it may be tempting to use ICLR points as a sole criterion for faculty cases. However, we believe that such practice is not appropriate as it does not account for the diverse experiences and contributions of candidates.

As Goodhart's law states, "when a measure becomes a target, it ceases to be a good measure." We believe that the use of ICLR points as a target may lead to unintended consequences. For instance, faculties may shift their publications to areas with higher ICLR points, which twists the original intention of the metric.

## A ETHICS

We believe that the benefits of our research significantly outweigh potential harms. While the research consists of analyzing publication records of faculties, such data is publicly available and widely used in academic hiring, promotion and tenure cases. The schedule of job talk candidates is also publicly available on the department website. Out of an abundance of caution, we erased all identifiable information of candidates in the paper to protect their privacy. We do not believe that our research will cause any tangible harm to the candidates or faculties in the dataset.

## B FULL DATA OF AREAS

In Table 3 and Table 4, we give the full data of areas and their conference publications and number of faculties.

**Table 3: Computer science areas and their corresponding top-tier conferences. The list of conferences is based on the CSRankings website [1].**

| Parent Area | Abbreviation | Area | Conferences |
|---|---|---|---|
| AI | ai | Artificial intelligence | AAAI, IJCAI |
| | vision | Computer vision | CVPR, ECCV, ICCV |
| | mlmining | Machine learning | ICLR, ICML, NeurIPS |
| | nlp | Natural language processing | ACL, EMNLP, NAACL |
| | inforet | The Web & information retrieval | SIGIR, WWW |
| Systems | arch | Computer architecture | ASPLOS, ISCA, MICRO |
| | comm | Computer networks | SIGCOMM, NSDI |
| | sec | Computer security | CCS, IEEE S&P ("Oakland"), USENIX Security |
| | mod | Databases | SIGMOD, VLDB |
| | da | Design automation | DAC, ICCAD |
| | bed | Embedded & real-time systems | EMSOFT, RTAS, RTSS |
| | hpc | High-performance computing | HPDC, ICS, SC |
| | mobile | Mobile computing | MobiCom, MobiSys, SenSys |
| | metrics | Measurement & perf. analysis | IMC, SIGMETRICS |
| | ops | Operating systems | OSDI, SOSP |
| | plan | Programming languages | PLDI, POPL |
| | soft | Software engineering | FSE, ICSE |
| Theory | act | Algorithms & complexity | FOCS, SODA, STOC |
| | crypt | Cryptography | CRYPTO, EuroCrypt |
| | log | Logic & verification | CAV, LICS |
| Interdisciplinary Areas | bio | Comp. bio & bioinformatics | ISMB, RECOMB |
| | graph | Computer graphics | SIGGRAPH, SIGGRAPH Asia |
| | csed | Computer science education | SIGCSE |
| | ecom | Economics & computation | EC, WINE |
| | chi | Human-computer interaction | CHI, UbiComp / Pervasive / IMWUT, UIST |
| | robotics | Robotics | ICRA, IROS, RSS |
| | visualization | Visualization | VIS, VR |

**Table 4: Full data of areas with faculty and publication metrics. This data reflects the time period of 5 years during 2019–2023.**

| Area | Faculty Count | Publication Count | Faculties Per Publication | ICLR Points |
|---|---|---|---|---|
| Algorithms & complexity | 483.34 | 2121 | 0.23 | 3.03 |
| Artificial intelligence | 1654.66 | 11662 | 0.14 | 1.89 |
| Computer architecture | 304.54 | 1292 | 0.24 | 3.14 |
| Embedded & real-time systems | 126.93 | 467 | 0.27 | 3.62 |
| Comp. bio & bioinformatics | 101.17 | 252 | 0.40 | 5.34 |
| Human-computer interaction | 989.32 | 5198 | 0.19 | 2.53 |
| Computer networks | 166.90 | 613 | 0.27 | 3.62 |
| Cryptography | 171.96 | 923 | 0.19 | 2.48 |
| Computer science education | 206.95 | 815 | 0.25 | 3.38 |
| Design automation | 303.00 | 2240 | 0.14 | 1.80 |
| Economics & computation | 60.40 | 639 | 0.09 | 1.26 |
| Computer graphics | 198.64 | 1473 | 0.13 | 1.80 |
| High-performance computing | 245.06 | 922 | 0.27 | 3.54 |
| The Web & information retrieval | 512.97 | 2854 | 0.18 | 2.39 |
| Logic & verification | 229.65 | 694 | 0.33 | 4.41 |
| Measurement & perf. analysis | 158.04 | 513 | 0.31 | 4.10 |
| Machine learning | 1716.55 | 22851 | 0.08 | 1.00 |
| Mobile computing | 190.81 | 917 | 0.21 | 2.77 |
| Databases | 484.64 | 2256 | 0.21 | 2.86 |
| Natural language processing | 766.99 | 9420 | 0.08 | 1.08 |
| Operating systems | 110.12 | 340 | 0.32 | 4.31 |
| Programming languages | 278.34 | 736 | 0.38 | 5.03 |
| Robotics | 757.93 | 11708 | 0.06 | 0.86 |
| Computer security | 655.63 | 2769 | 0.24 | 3.15 |
| Software engineering | 446.97 | 1455 | 0.31 | 4.09 |
| Computer vision | 1045.84 | 16751 | 0.06 | 0.83 |
| Visualization | 315.67 | 1333 | 0.24 | 3.15 |